# Online networks and subjective well-being [1]

Fabio Sabatini [2] [*]

Francesco Sarracino [3]

## Abstract

We argue that the use of online networks may threaten subjective well-being in several ways, due to the inherent attributes of Internet-mediated interaction and through its effects on social trust and sociability. We test our hypotheses on a representative sample of the Italian population. We find a significantly negative correlation between online networking and well-being. This result is partially confirmed after accounting for endogeneity. We explore the direct and indirect effects of the use of social networking sites (SNS) on well-being in a SEM analysis. We find that online networking plays a positive role in subjective well-being through its impact on physical interactions, whereas SNS use is associated with lower social trust. The overall effect of networking on individual welfare is significantly negative.

[1] We are grateful to participants at the Conference "Social and cultural changes in comparative prospect: values and modernization" (Moscow, March 29-A pril 6, 2014), and at seminars at the GESIS - Leibniz Institute for the Social Sciences in Cologne, at the Institut national de la statistique et des études économiques du Grand-Duché du Luxembourg, at the University of Milan Bicocca and at the Italian National Research Council for helpful comments. Usual caveats apply.

[2] Department of Economics and Law, Sapienza University of Rome, Italy, and Laboratory for Comparative Social Research (LCSR), National Research University Higher School of Economics, Russia (LCSR Russian Government Grant # 11.G34.31.0024 from November 28, 2010).

[*] Corresponding author. Postal address: Sapienza Università di Roma, Facoltà di Economia, via del Castro Laurenziano 9, 00161, Roma, Italy. E-mail: fabio.sabatini@uniroma1.it. Phone and fax: +39 0649766949.

[3] Institut National de la Statistique et des Études Économiques du Grand-Duché du Luxembourg (STATEC), Laboratory for Comparative Social Research (LCSR), National Research University Higher School of Economics, Russia (LCSR Russian Government rant # 11.G34.31.0024 from November 28, 2010) and GESIS Leibniz-Institute for the Social Sciences. Email: f.sarracino@gmail.com.

# 1. Introduction

Understanding the impact of online networks on people's well-being is a fundamental task for social research, given the dramatic rise in the use of social networking sites (SNSs) that has been registered in the last decade. This issue has so far been analyzed in narrow and biased samples, in most cases composed of non-representative groups of undergraduate students. This body of research has so far produced conflicting results, which account for endogeneity bias only limitedly.

We study the systemic effects that social networking sites (SNSs) may have on individual welfare in a bigger and nationally representative sample. We argue that online networks may threaten subjective well-being through three main mechanisms: SNSs use can affect users' welfare directly, because of the inherent attributes of online social interaction, and indirectly, through its effects on social trust and sociability.

Several authors have documented the association of heavy SNSs use with addiction and dependency (Young, 2004; Kim and Haridakis, 2014), anxiety (McCord et al., 2014), depression (Kraut et al., 2002) and other specific disorders such as bipolar-mania, narcissism and histrionic personality (Rosen et al., 2013). The underlying thesis in these studies is that online participation may per se be a detrimental factor of psychological well-being.

In addition, we believe there may be further ways in which online networking can affect individual welfare. SNSs use in fact entails a higher exposure to cultural diversity than face-to-face interactions. Online networks continuously create rooms for discussion where users "meet" strangers belonging to different socio-economic and cultural backgrounds. Differently from face to face interactions, where we usually select the people with whom to exchange ideas, selection mechanisms are inherently weak in online networks. For example, tolerant users may easily find themselves to interact with unknown, racist or homophobic readers in a "public" page. Think of the unexpected and unwanted interactions that take place on the Facebook page of a national newspaper, where a wide and heterogeneous audience can comment on news and editorials. Diversity is much more diffused in the global population of Internet users than in their limited reference groups. In principle, exposure to diversity might be considered as a source of knowledge and dialogue. However, the empirical literature has shown how interaction with heterogeneous preference types can also turn into a powerful source of frustration and distrust. In addition, interacting online entails a higher risk of being targeted with offensive behaviors and hate speech. This risk is particularly significant for women and users belonging to minorities or discriminated groups. Internet-mediated interaction often violates well-established face-to-face social norms for the polite expression of opposing views. In online discussions with unknown others, individuals more easily indulge in aggressive and disrespectful behaviors. Online networks also are a fertile ground for spreading harmful, offensive, or controversial contents often lying at the verge between free speech and hate speech.

As we will explain in Section 2, diversity and hate speech may have relevant effects on Internet users' trust in others, which in turn might affect people's well-being. Social trust has, in fact, been found to be one of the strongest predictor of self-reported happiness by empirical studies (Bjørnskov, 2003; Bruni and Porta, 2007).

Finally, it is not clear whether online networking supports or crowds-out face-to-face interactions. So far we have conflicting evidence between studies suggesting that SNSs can be an isolating medium leading to loneliness and less social interactions, and studies claiming that online networking actually supports physical interactions and deter from loneliness (Steinfield et al., 2008; Heliwell and Huang, 2013; Sabatini and Sarracino, 2014). We argue that a negative effect of SNSs on social interactions and trust would be a channel of further worsening in subjective well-being. A positive effect of networking on social interactions and trust, on the other hand, would be a factor increasing well-being that may to a certain extent counterbalance the negative effects on trust.

We empirically test our hypotheses on a nationally and regionally representative sample of the Italian population. We use pooled cross-sectional data including the two latest waves of the Multipurpose Survey on Households (MHS) provided by the Italian National Institute of Statistics (Istat). This survey contains detailed information on Internet use - with specific regard to participation in online networks - and the different dimensions of social capital. We use ordered probit models to check the partial correlations among our variables of interest. Differently from previous studies in the field, we also propose an empirical strategy to identify  the direct and indirect effects of SNSs on well-being. Endogeneity in online networking is addressed by exploiting technological characteristics of the pre-existing voice telecommunication infrastructures that exogenously determined the availability of broadband for high-speed Internet. We then disentangle the direct effect of SNS use on well-being from the indirect effect possibly caused by the SNSs' impact on trust and sociability through a structural equations model (SEM).

Our study innovates the literature in four substantive ways. We are the first to investigate the relationship between online networks, well-being, and social capital drawing on a large and nationally representative sample. Second, we contribute to the happiness economics literature with the first study on the role of online networks. Third, we carefully account for the endogeneity issues inherently related to this kind of analysis. We exploit the exogenous variation in individuals' access to fast Internet in their area of residence caused by the orographic heterogeneity of the Italian territory, to control for endogeneity bias through instrumental variables (IV) estimates. Fourth, we propose an explanation of the causal mechanism through which online networking may affect individual welfare. We focus on the direct effect of SNSs on psychological well-being and on the indirect effects mediated by networking's influence on social trust and sociability. We test this explanation empirically using Structural Equation Modelling.

The paper is organized as follows: in Section 2, we briefly review the literature and present our theoretical hypotheses. Section 3 describes our empirical strategy. Section 4 presents and discusses the empirical findings. The conclusion summarizes some lessons on the role of online networking in subjective well-being.

## 2. Related literature and theoretical hypotheses

The achievement of better living conditions is now one of the main objectives of political institutions. This is the result of about forty years of scientific research in the field of quality of life, which has gone far beyond the boundaries of the academic debate to involve mass media and the public opinion. People's evaluation of their own well-being, for short also referred to as "happiness" or "life satisfaction", is monitored through survey questions such as: "Taking all things together, how happy would you say you are?" or "All things considered, how satisfied are you with your life as a whole these days?". A consolidated literature proved the reliability of these subjective well-being measures (van Praag et al., 2003), which have then been employed in many fields of social research. This body of studies explored various determinants of individual well-being. Among these, social capital has been identified as a particularly strong predictor (Bjørnskov, 2003; Bartolini et al., 2013; Becchetti et al., 2008; Bruni and Stanca, 2008; Heliwell and Huang, 2009).

### 2.1 Online networks and individual well-being

A relatively new field of research explores how online participation and networking may influence Internet users' life satisfaction. In one of the seminal papers on the topic, Steinfield et al. (2008) proposed suggestive hints about the role that SNSs may play in reducing inequalities in well-being. The authors find that life-satisfaction and self-esteem serve to moderate the relationship between the intensity of Facebook usage and bridging social capital in a sample of undergraduate students at a Midwestern university. Those with lower self-esteem and those who are less satisfied with their life gained more from their use of Facebook in terms of bridging social capital than participants with higher self-esteem. In the authors' approach, the effect of online networking is basically mediated by its ability to deepen and expand users' social relationships. A few pioneering economic studies analysed the impact of broadband and, more in general, of Internet use. Pénard and Poussing (2010) found ambiguous results on the relationship between online investments in social capital and the development of face-to-face interactions among Luxembourg Internet users. In a following study, the authors found that non-users are less satisfied with their life than Internet users (Pénard and Poussing 2013).

Studies advancing less optimistic claims on the role of SNSs, on the other hand, highlighted that individual well-being may be directly affected by online networking due to the inherent characteristics of Internet-mediated interaction, which may cause addiction and disorders.

In a recent paper Kross et al. (2013) used experience sampling to show that Facebook use predicts negative shifts on subjective well-being over time in a group of 82 selected users. The more people used Facebook at one time point, the worse they felt afterwards; the more they used Facebook over two-weeks, the more their life satisfaction levels declined over time. The effects found by the authors were not moderated by the size of people's Facebook networks, their perceived supportiveness, motivation for using Facebook, gender, loneliness, self-esteem, or depression, thus suggesting the existence of a direct link between SNSs' use and subjective well-being. Kross and colleagues do not deepen the possible causal mechanism leading to a reduction in the subjective well-being of SNSs' users. However, their findings are consistent with a series of previous studies claiming the existence of statistically significant associations between heavy Internet or Facebook use and negative states of mind such as depression, anxiety, narcissism, bipolar mania, and other disorders. Other possible channels of decreasing well-being could be jealousy and envy. Based on responses to an online survey completed by 308 undergraduate students, Muise et al. (2009) concluded that increased Facebook exposure predicts jealousy above and beyond personal and relationship factors. Krasnova et al. (2013) used responses from 584 Facebook users recruited via email to show that passive following exacerbates envy feelings, which decrease life satisfaction. The authors noted that about one-fifth of all recent events that had provoked envy among respondents took place within a Facebook context. This reveals a prominent role of the platform in users' emotional life.

Consistently with the studies mentioned above, we assume the following hypothesis:

**Hypothesis 1**: SNSs' use worsens subjective well-being.

The effect of SNSs on SWB may also be mediated by omitted phenomena we are not able to control for, such as envy or depression. More specifically, we propose that the influence of networking on well-being might depend on how Web-mediated social participation affects trust and sociability. We briefly explain these issues in the following subsections.

## 2.2 Online networks, trust, and individual well-being

To trust is to assume that a person or institution will "observe the rules of the game" (Citrin and Muste 1999: p. 465) and to believe that those involved will act "as they should" (Barber 1983). In the words of Mutz and Reeves (2005: p. 3): "In face-to-face settings where people disagree about politics, there are strong social norms likely to be observed for purposes of these interactions. Face-to-face exchanges are relatively polite. Although people occasionally yell at one another and stomp their feet over political differences, such behavior is far more common in mediated presentations of political views. Norms involving politeness are extremely strong; most people are polite most of the time."

In principle, people might expect SNSs users to behave in online networks by the same social norms usually acknowledged in physical interactions. However, as reported in Sabatini and Sarracino (2014), SNSs offer public forums for discussion – such as Facebook's "public pages", "groups", and "communities", as well as commenting platforms for online magazines and newspapers (e.g. Disqus) – where individuals are likely to deal with strangers in a more aggressive and unscrupulous way than they would in a physical meeting. In online environments, unknown strangers are "invisible" and their reaction to provocative behaviors may be easily neutralized (for example by simply withdrawing from the conversation, or even by "blocking" them through the network's privacy settings). As a result, people care less of the risk of offending others in a conversation. In physical interactions, we usually think twice before insulting a person who politely expresses an opposing view. In online interactions, dealing with strangers who advance opposite views in an aggressive and insulting way seems to be a widespread practice, whatever the topic of discussion is.

Following the argument of Mutz and Reeves (2005) about televised incivility, we hypothesize that when unknown others violate interpersonal social norms and behave aggressively and offensively in online environments, people react as they would if those aggressions and offenses were perpetrated in real life. We argue that this mechanism may cause anxiety, distress, and deterioration in trust towards unknown others.

The use of online networks may destroy trust even in individuals who are not targeted with hate speech or aggressive behaviors. In fact, compared to face-to-face interactions, online networks allow users to silently observe the opinions and behaviors of an immensely wider share of their fellow citizens. The psychological literature has shown that most people tend to overestimate the extent to which their beliefs or opinions are typical of those of others. There is a tendency for people to assume that their own opinions, beliefs, preferences, values, and habits are "normal" and that others also think the same way that they do. This cognitive bias leads to the perception of a consensus that does not exist, or a "false consensus" (Gamba, 2013). This bias is especially prevalent in closed networks where individuals tend to think that the collective opinion of their reference group matches that of the larger population. Online networks, however, allow users to acquire more detailed information about the preference types and behaviors spread outside of individuals' narrow reference groups. People who previously felt part of a majority may discover to be surrounded with preference types they dislike (e.g. a racist person may find out that most people appreciate ethnic diversity, or vice versa), and this might lead to a revision of individuals' beliefs about the trustworthiness of others.

Both the mechanisms illustrated above lead to hypothesize that the use of SNSs might reduce trust, as it has been recently documented in Sabatini and Sarracino (2014).

The reduction in trust may thus be an important mediator of the relationship between online networking and subjective well-being.

**Hypothesis 2a**: The erosion of social trust negatively correlates with subjective well-being.

**Hypothesis 2b**: The relationship between SNSs' use and subjective well-being is mediated by the negative effect of SNSs' use on social trust.

## 2.3 Online networks, sociability, and individual well-being

Early sociological studies on Internet use shared the concern that the Internet would crowd out social interaction. The main argument of this body of research was based on the presumption that the more time people spend using the Internet during leisure time, the more time has to be detracted from face-to-face social activities (e.g. Nie et al. 2002, Gershuny 2003). These studies, however, date back to shortly before the explosion of online networking, and they could not differentiate between pure entertainment and social activities. More recent studies found that SNSs strengthen bonding and bridging social capital (Steinfield et al. 2008, Park et al. 2009), allow the crystallization of weak or latent ties that might otherwise remain ephemeral (Haythornthwaite 2005, Ellison et al. 2007: 2011), support teenagers' self-esteem - encouraging them to relate to their peers (Ellison et al. 2007; 2011; Steinfield et al. 2008), enhance civic engagement (Zhang et al. 2010) and political participation (Gil de Zúñiga et al. 2012).

Participation through online networks can help individuals to maintain their social contacts from distant locations. In addition, Internet-mediated interaction is less sensitive to the reduction in leisure time caused by an intense pace of work. As argued by Antoci et al. (2012), online social participation favors asynchronous interactions that allow individuals to compensate for the lack of time: one can benefit from the others' participation, for example by reading a message or a note, even if the person who wrote it is currently offline.

Some SNSs serve the explicit purpose of creating new bridges between members having common interests and beliefs. In this paper we test the hypothesis that online networking has a positive indirect effect on subjective well-being that is mediated by face-to-face interactions.

**Hypothesis 3a**: Face to face social interactions positively correlate with subjective well-being.

**Hypothesis 3b**: The relationship between SNSs' use and subjective well-being is mediated by the positive effect of SNSs' use on face to face interactions.

## 2. Data and Methodology

We use a pooled cross-section of data drawn from the last two waves (2010 and 2011) of the "Multipurpose Survey on Households" (MHS) provided by the Italian National Institute of Statistics (Istat). This survey uses face-to-face interviews to investigate a wide range of social behaviors and perceptions on a nationally and regionally representative sample of approximately 24,000 households, roughly corresponding to 50,000 individuals. Subjective well-being is observed through the answers to the question "How satisfied are you with your life as a whole nowadays?". Answers range on a scale from 0 (extremely dissatisfied) to 10 (extremely satisfied), which is a widely adopted scale for measuring well-being (Pavot and Diener, 1993; Krueger and Schkade, 2008).

The reliability of these measures has been corroborated by experimental evidence from several disciplines. For example, subjective well-being correlates with objective measures of well-being such as the heart rate, blood pressure, frequency of Duchenne smiles and neurological tests of brain activity (Blanchflower and Oswald, 2004; van Reekum et al., 2007). Moreover, subjective measures of well-being are strongly correlated with other proxies of subjective well-being (Schwarz and Strack, 1999; Wanous and Hudy, 2001; Schimmack et al., 2010) and with judgements about the respondent's happiness provided by friends, relatives or clinical experts (Kahneman and Krueger, 2006; Layard, 2005).

[Figure 1 here]

Figure 1 compares the distribution of life satisfaction in 2010 and 2011 across Italian regions. Two main features arise: first, well-being is on average higher in northern regions, varying between 7.25 and 7.76 in the North, 7.02 and 7.21 in the Center and 6.85 and 7.26 in the South; second, between 2010 and 2011 the average level of life satisfaction slightly decreased in almost every region.

We observe social capital through indicators of its structural and cognitive dimensions. The structural dimension is given by social interactions, as measured by the frequency of meetings with friends. Respondents were asked to report how many times they meet their friends on a scale from 1 (in case they have no friends) to 7 (if respondents meet their friends every day). Cognitive social capital is given by social trust, as measured by binary responses to the question: "Do you think that most people can be trusted, or that you can't be too careful in dealing with people?" as developed by Rosenberg (1956).

We also use a further indicator of social trust drawn from the so-called "wallet question" to check the robustness of our findings. The wording is as follows: "Imagine you lost your wallet with your money, identification or address in your city/area and it was found by someone else. How likely do you think your wallet would be returned to you if it were found by a neighbor/the police/a stranger?" Possible responses were: "Very likely", "Fairly likely", "Not much likely", and "Not likely at all". The data on the frequency of wallet returns were later used by Knack (2001) to provide some behavioral validation for the use of answers to the "Rosenberg question" on generalized trust. Knack (2001) found that at the national level the actual frequency of the returns correlated at the 0.65 ($p < 0.01$) level with national average responses to the general social or interpersonal trust question (as measured by the World Values Survey). Here we followed Knack (2001) and measured social trust based on the responses to the hypothesis that the wallet was found by a complete stranger. We reversed the scale, so that larger values indicate greater trust in unknown others.

Online social interactions are observed by means of a dichotomous variable capturing respondents' participation in SNS's such as Facebook and Twitter. To explore the relationship among subjective well-being, actual and virtual social capital, we adopted an ordered probit model with robust standard errors. Hence, if subjective well-being is ordered in 11 categories, then the resulting model is:

$$SWB_i = \begin{cases} 1 & \text{if } y_i \leq 0 \\ 2 & \text{if } 0 < y_i \leq c_1 \\ 3 & \text{if } c_1 < y_i \leq c_2 \\ & . \\ & . \\ & . \\ 10 & \text{if } c_{10} < y_i \end{cases} \qquad (1)$$

Where $0 < c_1 < c_2 < ... < c_{10}$; the index $\boldsymbol{i}$ stands for individuals;

$SWB_i = \alpha + \beta_1 \cdot friends_i + \beta_2 \cdot trust_i + \beta_3 \cdot fb_i + \theta \cdot X_i + \epsilon_i, \epsilon_i \backsim N(0,1);$

And $c_{10}$ are unknown parameters to be estimated.

The list of control variables ($X$) includes the kind of technology available to the respondent to connect to the Internet along with individual's age (both in linear and squared form), gender, marital status, number of children, education, work status, and the time spent in commuting and watching television (in minutes).

Table 1 provides a summary of descriptive statistics.

[Table 1 here]

## 2.1 Controlling for endogeneity

The coefficients from equation 1 indicate the sign and magnitude of partial correlations among variables. However, we cannot discard the hypothesis that our main explanatory variables are endogenous to subjective well-being. In particular, while the effect on well-being has been largely explored in the case of face-to-face interactions, we do not have conclusive evidence about the endogenous relationship between online interactions and well-being. This weakness suggests caution about the generality of the results provided by previous literature. Individual effects such as personal characteristics may be correlated with both participation in SNSs and well-being. Happier people may also be more outgoing and open-minded, and may have a higher propensity for various kinds of social interaction. The inclusion of a wide set of control variables is intended to reduce the possible influence of omitted variables both at the individual and at the local level.

However, this is not enough to avoid the possible bias induced by reverse causality. For example, happy people may want to share their feelings or information about positive life events on online networks with their important persons. This is why just-married couples often upload pictures of their wedding ceremony on Facebook. On the other hand, lonely and/or unhappy individuals may want to use Facebook with the hope of improving their condition by establishing new relationships and sharing their feelings. For example, divorced people - who usually report severely low levels of happiness - may want to use online networking to find new mates and start a romantic relationship.

To deal with this problem, we turn to instrumental variables estimates using a two stage least squares (2SLS) model (Wooldridge, 2002) where, in the first stage, we instrument our two measures of online networking.

We adopted two instruments that can be easily shown to be exogenous to subjective well-being (our dependent variable) – orthogonality condition -- and not driven by individuals' propensity for online networking (the main endogenous variable) – relevance condition: 1) The share of the population for whom a DSL connection was available in respondents' region of residence. DSL ("digital subscriber line", originally "digital subscriber loop") is a family of technologies that offers access to the Internet by exchanging digital data over the wires of a telephone network. Data are retrieved from the Italian Ministry of Economic Development. 2) The percentage of the region's area that is not covered by optical fiber, which represents a measure of digital divide. Optical fiber allows exchange of information over long distances and at higher bandwidths (data rates) than DSL, thus providing a fast Internet connection. Data are based on figures from the Italian Observatory on Broadband.

Both instruments were observed in 2008, two years before the first of the two waves of the Multipurpose Household Survey were collected. There are various reasons to believe that the 2008 level of regional DSL coverage is not *directly* correlated to the individual level of subjective well-being in the period 2010-11. The availability of DSL is a pre-condition for the individual choice to purchase a high-speed access that may create room for the development of online interactions, which in turn may influence individual welfare in a variety of ways. Hence, we assume that the effect of broadband coverage on subjective well-being (and social capital) occurs through the use of social networking sites, chats, forums, newsgroups and similar forms of web-mediated communication.

Our assumption that the differences in the availability of DSL are exogenous to subjective well-being is derived from the environmental features of the Italian territory, which have played (and currently play) a major role in the development of Italy's infrastructures for accessing fast Internet. DSL technology is based on the transmission of data over the user's copper telephone line, i.e. over pre-existing voice telecommunications infrastructures. However, the availability of a telephone infrastructure is a necessary, but not sufficient, condition for the availability of broadband. What really matters is the so-called "local loop", i.e. the distance between final users' telephone line and the closest telecommunication exchange or "central office" (Grubesic, 2008; OECD, 2009; Czernich, 2012; Campante et al., 2013). For the supply of traditional voice services, the length of this distance does not affect the quality of the connection. This is why, before the advent of the Internet, the former state monopoly phone carrier (*Telecom Italia*) did not pay any attention to local loops, whose length was entirely determined in accordance to the orographic features of the territory. However, this distance matters for the provision of fast Internet because the longer is the copper wire, the less bandwidth is available via this wire. In particular, if the distance is beyond a threshold of approximately 4.2 kilometers (about 2.61 miles), then the band of the copper wires is not wide enough to allow a fast Internet connection (Grubesic, 2008; Czernich, 2012). In this case, it is impossible to implement the broadband connection through traditional copper wires. This is the case, for example, of Italian rural areas, which represent more than half of the Italian territory and comprise severely isolated and less densely populated highlands or hills. In 2007, a large part of these areas were characterized by a high length (≥ 4.2 kilometers) of local loops, which ultimately is the result of the imperviousness of the territory. Therefore, in most cases, these areas lacked the necessary infrastructures for the diffusion of the DSL broadband (Ciapanna and Sabbatini, 2008; Agcom, 2011). Figure 2 presents a map of the broadband coverage of the Italian territory, in comparison with its orographic characteristics.

[Figure 2 here]

Hence, self-reported happiness in 2010-11 is not correlated with the distribution of DSL infrastructures in 2008 because the latter strictly depends on local loops, whose location was determined many years before the rise of the Internet and based on the orographic features of the territory.

The arguments supporting the assumption of the orthogonality of the share of the population covered by DSL are compelling for the second instrument. When, as explained above, the broadband connection cannot be implemented through pre-existing copper wires, it is necessary to turn to an optical fiber-based technology. The possibility and the costs of installing this type of infrastructure, however, even more strongly rely on the exogenous characteristics of the natural environment. Differently from DSL, in fact, optical fiber entails the need to install new cables underground. This involves excavation projects, which are much more costly and generally delay or even prohibit the provision of broadband in the area. Once again, orographic differences between regions must be considered as a "natural" cause of the digital divide which generated a variation in access to fast Internet across regions that is exogenous to people's well-being and cannot be driven by their preference for online networking. The assumption of orthogonality of the instruments is confirmed by the tests of over-identifying restrictions we run in the context of IV estimates (reported in Section 3).

For any given set of orographic characteristics of the area, the provision of broadband – whether through DSL or optical fiber technology – may also have been influenced by some socio-demographic factors that affected the expected commercial return on the provider's investment, such as population density, per capita income, the median level of education and the local endowments of social capital. To account for the eventual confounding effects of these features, we included regional GDP per capita in real euros of 2005, along with a set of regional and year fixed effects. However, we emphasize that an eventual correlation between the commercial return of the investments in fast Internet connections with well-being, on one side, and the instruments, on the other, does not raise any concern of confounding the causal interpretation. The reason is that the instruments do not determine the confounders, thus excluding the hypothesis of indirect causal mechanisms.

To perform 2SLS estimates with a dichotomous endogenous variable and a categorical dependent one, we used a multi-equation conditional mixed-process estimator as implemented by Roodman (2011). This technique allows us to adopt a probit model to estimate the first step regression where SNS is regressed over the two instruments (and the control variables) and an ordered probit model to fit the second step where the dependent variable is life satisfaction.

Formally, the first step of the 2SLS model can be written as:

$$fb_i = \begin{cases} 0 & \text{if } y_i \leq 0 \\ 1 & \text{if } y_i > 0 \end{cases} \qquad (2)$$

Where $fb_i = \pi_1 + \pi_2 \cdot z_1 + \pi_3 \cdot z_2 + \pi_4 \cdot X_i + v_i, v_i \backsim N(0,1)$ and $z_1$ and $z_2$ are the two above-mentioned instruments.

The model of the second step is as follows:

$$SWB_i = \begin{cases} 1 & \text{if } y_i \leq 0 \\ 2 & \text{if } 0 < y_i \leq c_1 \\ 3 & \text{if } c_1 < y_i \leq c_2 \\ & \cdot \\ & \cdot \\ & \cdot \\ 10 & \text{if } c_{10} < y_i \end{cases} \qquad (3)$$

Where $0 < c_1 < c_2 < ... < c_{10}$; the index $i$ stands for individuals;

$SWB_i = \alpha + \beta_1 \cdot friends_i + \beta_2 \cdot trust_i + \beta_3 \cdot \widehat{fb_i} + \theta \cdot X_i + \epsilon_i, \epsilon_i \backsim N(0,1)$;

And $c_{10}$ are unknown parameters to be estimated.

$\widehat{fb_i}$ is the predicted probability of using SNS from the first step and $c_{10}$ are unknown parameters to be estimated.

As in model 1 SWB is measured through the life satisfaction question; $\theta$ is a vector of parameters of the control variables X; $\beta_3$ is the coefficient of the use of SNS; $\widehat{fb_i}$ is the instrumented use of SNS and $\epsilon_i$ is the error term.

To further check the robustness of our estimates, we also test the relationship among our variables using a linear 2SLS model[4].

Finally, to test for possible indirect effects of SNS on subjective well-being through actual social capital, we tested the following structural equation model.

---

[4] Results are available upon request to the authors.

[Figure 3 here]

### 3 Partial correlations using ordered probit

We first report, in model 1, how the covariates correlate with the dependent variable. Life satisfaction is found to be significantly and negatively correlated with the time spent watching television. This result is consistent with previous studies analyzing the effect of television on individual happiness (Frey et al., 2007; Bruni and Stanca, 2008). The relationship between life satisfaction and age follows an inverted-U shape curve. This result suggests that people's well-being decreases with age up to a minimum that, in our sample, corresponds to about 30 years. Afterwards, the relationship between well-being and age turns positive. This result is consistent with previous findings of the economic literature about the relationship between aging and well-being (Blanchflower and Oswald, 2008).

We also controlled for the kind of connection used by individuals to connect to the Internet (e.g. modem, DSL, fiber, satellite, etc.). As expected, none of them was found to have a statistically significant relationship with life satisfaction. Other socio-demographic controls, such as education, marital and work status have all the expected signs and are omitted from tables for the sake of brevity.

[Table 2 here]

In model 2, we introduced the frequency of meetings with friends and social trust. Both variables are significantly and positively correlated with life satisfaction. This result is consistent with Hypotheses 2b and 3b and with previous literature examining the role of relational goods in individual happiness (Becchetti et al., 2008; Sarracino, 2010; Bartolini et al., 2013). Friendships can improve life satisfaction in a number of ways, from the provision of social support in case of need to the pleasure of spending time together.

The significant and positive coefficient of social trust, on the other hand, is consistent with empirical studies claiming the existence of a link between various forms of trust and life satisfaction across countries (Bjørnskov, 2003) and at individual level (Helliwell, 2003; Helliwell et al., 2009; Helliwell and Wang, 2011). For example, Bjørnskov (2003) suggested that the cross-country relationship between trust and well-being may be due to the higher economic growth rates generally connected to higher levels of social trust (see, for example, Knack and Keefer, 1997). In addition, social trust could help countries to successfully cope with external shocks, as suggested by recent studies on Japanese earthquakes (Yamamura, 2014). The ability to cope successfully with external shocks could also help promote stability in the economy. This in turn may reduce economic uncertainty, further benefiting life satisfaction.

As expected, there is a significant and positive relationship between self-reported health and life satisfaction (the sign of the coefficients reported in Table 2 is negative because higher values of the health indicator correspond to worse self-reported well-being), consistently with previous research on Italy (Sabatini, 2014).

Models 4 to 6 show that there is a weakly significant and negative correlation between online networking and subjective well-being. This result is per se interesting and it provides preliminary support to Hypothesis 1, but it must be handled with caution due to the sources of potential endogeneity we described in the previous sections. On the one hand, the negative correlation supports skeptical views suggesting that devoting too much time to online networking may undermine life satisfaction (e.g. Kross et al., 2013).

The result about online networking reported in Table 2, however, may be caused by the fact that individuals who are socially anxious and less satisfied with their life are more likely to use online networks to reduce their loneliness. The negative relationship between online networking and subjective well-being may thus be due to the negative feelings potentially associated with higher levels of online networking.

In addition, other confounding factors that we are not able to control for may bias both the dependent and the independent variables in our regressions. For example, the number of Facebook friends, the perceived supportiveness of users' online and physical social environment, the presence of depressive symptoms (such as low self-esteem) may also play a role. Hence we turn to instrumental variables (IV) described in section 2 to effectively tackle endogeneity issues.

## 4. Instrumenting the use of SNSs

Our IV approach uses the percentage of the population for whom DSL connection was available in respondents' area of residence in 2008 and the percentage of the region's area that was not covered by optical fiber in 2008 as instruments for the individual propensity for online networking in the period 2010-2011. Table 3 reports IV estimates of the determinants of life satisfaction. The statistical insignificance of online networking suggests that, even if participation in SNSs is correlated with lower satisfaction, it may hardly be considered as a cause of decreasing well-being per se. The first stage of IV estimates, along with the test of the joint significance of coefficients, confirms the relevance of instruments.

[Table 3 here]

[Table 4 here]

Online networking loses its significance, whereas social trust and the frequency of meetings with friends are confirmed to be strong correlates of life satisfaction. IV estimates suggest that the significant and negative correlation between online networking and life satisfaction found in ordered probit regressions was spurious. This finding questions the validity of Hypothesis 1 and provides support to Helliwell and Huang (2013), who found that "real-life social networks" positively contribute to self-reported happiness, while the size of online networks is not a relevant predictor of subjective well-being.

## 5. The indirect effect of SNSs' on well-being

In order to disentangle the drivers of this correlation, we estimated the structural equation model described in Figure 3 (page 19).

In this model, we simultaneously estimated the effect of online networking on subjective well-being, on social trust, and on the frequency of meetings with friends jointly with the effect of the latter two dimensions on happiness. This empirical strategy allows us to better understand whether online networking, per se, impacts life satisfaction, or if the effect on happiness is mediated by the impact of online participation on users' social capital.

[Table 5 here]

Modification indexes take all values lower than 3.84, thus confirming the goodness of current model. The estimates suggest that using SNS reduces trust in others by 2.7% and on average it increases the frequency of meetings with friends by 8.42%. Both social trust and the frequency of meetings with friends are in turn strongly and positively correlated with subjective well-being, as predicted by Hypotheses 2a and 3a. These results suggest that the use of SNSs may indirectly affect individual welfare in two opposite ways: negatively, through a reduction in social trust, and positively, through the support of face-to-face interactions, consistently with Hypotheses 2b and 3b. The indirect effect of SNS on well-being mediated by social trust is about -0.07%, whereas the effect mediated by meetings with friends is about 0.68%, with a total indirect effect of about 0.57%.

However, if we also account for the direct negative effect of SNS use on life satisfaction, then the total net effect is negative and amounts to about -0.15%. This result suggests that Hypothesis 1 is confirmed only if we keep in consideration the indirect effects of the use of SNSs on well-being. Goodness of fit measures are reported at the bottom of Table 5. Since the model chi-square is affected by sample size, following Kline (2005), we divide its value by the degrees of freedom of the model, obtaining the Normed Chi Square (NC). In general, as the sample size gets larger, the reliability of overall fit measures is reduced. In addition, it must be noted that values of fit indexes only indicate the average or overall fit of a model, and that it is possible that some parts of the model poorly fit the data even if the value of a particular index seems favorable (Kline, 2005). It thus seems reasonable in our case to focus on the significance of estimates' coefficients, which are also reported in Figure 4.

[Figure 4]

Overall, the SEM analysis reveals that the significantly negative correlation between online networking and subjective well-being may result from the combination of three main drivers:

1. An indirect positive effect due to the positive correlation between online networking and face-to-face interactions that in turn positively affect well-being, which is consistent with Hypotheses. 3a and 3b.

2. An indirect negative effect due to the negative correlation between online networking and social trust that in turn positively affects well-being, which is consistent with Hypotheses 2a and 2b.

3. A negative correlation between online networking and subjective well-being that is largely due to the indirect effects of SNSs on social trust and, therefore, on SWB. This evidence is consistent with Hypothesis 1.

The positive effect of online networking on face-to-face interactions is in line with previous findings from applied psychology (Steinfield et al., 2008) and economics (Becchetti and Degli Antoni, 2010). Apparently, social networking sites play a positive role in helping Internet users to preserve their relationships against the threats posed by busyness and distance.

The negative relationship of online networking with social trust, on the other hand, contradicts part of the previous literature on the topic. This may be due to the fact that empirical studies finding moderate and positive effects of Facebook use on trust in others commonly drew on very limited – and to a certain extent biased – samples, in most cases composed of small communities of undergraduate students enrolled in specific American colleges. As argued in Sabatini and Sarracino (2014), the "radius" of trust that college students may have in mind when responding to the trust question is likely to be relatively limited – and basically referred to their peers. In our study, we account for a large nationally and regionally representative sample of the Italian population, where the radius of trust is likely to be higher than that of students attending a specific college (see, for example Delhey et al., 2011).

The detrimental effect on trust in others may be interpreted as a consequence of users' interaction with unknown people on Facebook, Twitter, and commenting platforms such as Disqus. These platforms, in fact, create rooms for discussion in which selection mechanisms are weak or absent, differently from what happens in face-to-face interactions where we usually select a narrow circle of well-known friends and acquaintances to discuss political and moral issues. The Facebook page of a newspaper, for example, gathers a very heterogeneous audience who can comment on news and op-ed articles without moderation. Threads in these pages often allow the development of endless online discussions – that are generally encouraged by the pages' managers and by the platform itself, as they bring more visitors and "clicks" – in which individuals are forced to "meet" strangers, and often happen to encounter a wide variety of points of view. Empirical studies have shown that, at least in the short run, diversity along ethnic, religious, age, and socio-economic status lines may be a powerful source of frustration and distrust towards unknown others (Alesina and La Ferrara, 2002; Christoforou, 2011).

In online discussions with unknown others, individuals often exhibit a higher propensity for aggressive behavior than in face-to-face interactions. In addition, online conversations are more vulnerable to incomprehension and misunderstandings. In our Italian case study, the rising practice of hate speech, jointly with Facebook's increasing failures in identifying and removing it, suggests that unfriendly Internet-mediated communication with strangers may be an important channel of destruction of social trust (Sabatini and Sarracino, 2014).

Worries about hate speech have been recently stressed by the action of organizations advocating against gender-based discrimination (e.g. Women, Action and The Media, and The Everyday Sexism Project) and of groups which have historically faced discrimination in society that prompted a rethinking of Facebook's moderation policy.

On the other hand, SEM estimates also point out that SNS use has a net detrimental effect on well-being that is mediated by the negative effect that web-mediated interactions exert on trust.

## 6. Conclusions

In this paper, we carried out the first empirical analysis of the relationship between online networking and subjective well-being in a large and nationally representative sample. We first analyzed the correlation among variables using ordered probit models. We found the existence of a significantly negative correlation. We then addressed endogeneity in individuals' propensity for online networking by exploiting regional technological characteristics of the preexisting voice telephony network that exogenously determined the availability of broadband for accessing high-speed Internet.

When we addressed causality in IV estimates, the significance of the correlation between participation in social networking sites and subjective well-being disappeared. Ordered probit and IV estimates showed that face-to-face interactions and social trust are strongly and positively associated with well-being.

To disentangle the direct effect of SNS use from the change in well-being that may be caused by SNSs' impact on trust and sociability, we turned to a structural equations model. We found that online networking plays a positive role in subjective well-being through its impact on physical social interactions. On the other hand, SNS use is associated with lower social trust, which is in turn positively correlated with subjective well-being. The overall effect of networking on individual welfare identified by the structural equations model is significantly negative. These results are in line with Sabatini and Sarracino (2014), who found that participation in SNS might destroy social trust, and with Helliwell and Huang (2013), who found that face-to-face interactions are positively associated with happiness, while online networks are not.

The cross-sectional nature of the data employed in this study certainly suggests caution in the interpretation of findings, which may result from spurious correlations. However, the study contributes to the literature on Internet use and subjective well-being in a number of ways. This is the first empirical investigation of the relationship between Internet use and subjective well-being that explicitly accounted for the way in which the Internet is actually used, with a specific focus on social networking sites such as Facebook and Twitter. In addition, this is the first time that the role of online networks is addressed in a large nationally and regionally representative sample. Finally, this is the first time these issues have been addressed in a Mediterranean country.

The role of online networks in the development of interpersonal relationships and in the preservation of social cohesion suggests that individuals and communities who do not have access to the Internet – due, for example, to the absence of DSL or fiber infrastructures, or to lack of the skills required to participate in SNSs – may increasingly suffer from difficulties in social integration. From this point of view, the digital divide is likely to become an increasingly important factor of social exclusion, which may exacerbate inequalities in well-being and capabilities. A straightforward policy implication of this issue is that public institutions should ensure equal opportunities for connecting to fast Internet across regions (e.g. urban vs. rural), age cohorts, and social classes.

On the other hand, online networking exposes individuals to the risk of worsening people's trust in others and therefore people's life satisfaction. This finding suggests the need to update social networking sites' policies against hate speech and aggressive behaviors, as already requested by a growing number of advocacy groups, particularly focusing on gender- or race-based hate. In a note published on May 28, 2013 as a response to groups advocating against dis crimination and hate speech on social media, a Facebook manager stated that, even if the platform prohibits "Content deemed to be directly harmful", it intentionally allows "content that is offensive or controversial" with the aim of defending the principles of freedom of self-expression on which Facebook is founded. Harmful content is defined "as anything organising real world violence, theft, or property destruction, or that directly inflicts emotional distress on a specific private individual (e.g. bullying)", while no definition is provided for "offensive and controversial" content. To cope with hate speech issues, Facebook recently promised "to review and update guidelines, improve moderators' training, establish more formal lines of communication with advocacy groups and increase accountability of the creators of content which is cruel or insensitive but does not qualify as hate speech". The improvement in moderation may be looked as a key tool in fighting those behaviors that may cause a loss of trust by social networks' users.

# Tables

**Table 1:** Descriptive statistics

| variable | mean | sd | min | max | obs |
|----------|------|-----|-----|-----|-----|
| life satisfaction | 7.190 | 1.680 | 0 | 10 | 77560 |
| frequency of meetings with friends | 5.104 | 1.466 | 1 | 7 | 78988 |
| social trust | 0.223 | 0.416 | 0 | 1 | 77723 |
| wallet from stranger | 1.623 | 0.726 | 1 | 4 | 77368 |
| online networking | 0.453 | 0.498 | 0 | 1 | 35282 |
| women | 0.521 | 0.500 | 0 | 1 | 79433 |
| age | 50.11 | 18.21 | 18 | 90 | 79433 |
| age squared/100 | 28.43 | 19.07 | 3.240 | 81 | 79433 |
| minutes spent commuting | 18.67 | 12.32 | 0 | 57 | 36111 |
| minutes spent watching TV | 5.147 | 11.51 | 0 | 59 | 59924 |
| marital status | – | – | 1 | 4 | 79433 |
| educational status | – | – | 1 | 5 | 79433 |
| occupational status | – | – | 1 | 7 | 79433 |
| number of children | 1.011 | 1.009 | 0 | 7 | 79433 |
| real GDP per capita (thousands €2005) | 22.92 | 5.746 | 14.88 | 30.77 | 79433 |
| region | – | – | 10 | 200 | 79433 |
| year | – | – | 2010 | 2011 | 79433 |

**Table 2:** Relationship between SNSs and life satisfaction. Regressions with ordered probit.

| | (1) | (2) | (3) | (4) | (5) | (6) | (7) | (8) |
|---|---|---|---|---|---|---|---|---|
| life satisfaction | | | | | | | | |
| minutes spent commuting | -0.000892 | -0.000632 | -0.000670 | -0.000907 | -0.000691 | -0.000665 | -0.000704 | -0.000680 |
| | (-1.29) | (-0.94) | (-0.99) | (-1.29) | (-1.03) | (-0.99) | (-1.05) | (-1.01) |
| minutes spent watching TV | -0.00231*** | -0.00251*** | -0.00244*** | -0.00232*** | -0.00252*** | -0.00240*** | -0.00252*** | -0.00240*** |
| | (-3.97) | (-4.28) | (-4.15) | (-4.01) | (-4.37) | (-4.06) | (-4.36) | (-4.05) |
| women | 0.0158 | 0.0290 | 0.0237 | 0.0135 | 0.0273 | 0.0224 | 0.0271 | 0.0222 |
| | (0.89) | (1.62) | (1.32) | (0.75) | (1.47) | (1.23) | (1.47) | (1.22) |
| age | -0.0307*** | -0.0280*** | -0.0281*** | -0.0319*** | -0.0287*** | -0.0290*** | -0.0287*** | -0.0290*** |
| | (-7.32) | (-7.03) | (-6.95) | (-7.59) | (-7.23) | (-7.06) | (-7.25) | (-7.09) |
| age squared/100 | 0.0303*** | 0.0266*** | 0.0275*** | 0.0311*** | 0.0267*** | 0.0277*** | 0.0266*** | 0.027*** |
| | (5.90) | (5.49) | (5.57) | (6.13) | (5.51) | (5.51) | (5.53) | (5.53) |
| good health | -0.392*** | -0.384*** | -0.390*** | -0.393*** | -0.387*** | -0.391*** | -0.387*** | -0.390*** |
| | (-15.13) | (-14.83) | (-15.58) | (-15.18) | (-14.74) | (-15.87) | (-14.76) | (-15.89) |
| neither good nor bad health | -0.819*** | -0.795*** | -0.807*** | -0.821*** | -0.796*** | -0.806*** | -0.795*** | -0.805*** |
| | (-22.49) | (-20.93) | (-22.02) | (-22.40) | (-20.93) | (-22.34) | (-20.96) | (-22.36) |
| bad health | -1.187*** | -1.163*** | -1.176*** | -1.189*** | -1.168*** | -1.177*** | -1.167*** | -1.176*** |
| | (-12.43) | (-11.78) | (-11.63) | (-12.51) | (-11.84) | (-11.69) | (-11.84) | (-11.68) |
| very bad health | -0.912*** | -0.873*** | -0.896*** | -0.917*** | -0.880*** | -0.900*** | -0.879*** | -0.898*** |
| | (-3.82) | (-3.64) | (-3.74) | (-3.86) | (-3.70) | (-3.78) | (-3.68) | (-3.76) |
| modem | 0.0480 | 0.0393 | 0.0426 | 0.0404 | 0.0319 | 0.0381 | 0.0325 | 0.0388 |
| | (0.83) | (0.68) | (0.75) | (0.69) | (0.55) | (0.68) | (0.56) | (0.69) |
| dsl | -0.0126 | -0.0220 | -0.0203 | -0.0154 | -0.0249 | -0.0205 | -0.0241 | -0.0196 |
| | (-0.27) | (-0.47) | (-0.43) | (-0.33) | (-0.53) | (-0.43) | (-0.52) | (-0.42) |
| fiber | -0.0395 | -0.0515 | -0.0518 | -0.0438 | -0.0526 | -0.0479 | -0.0535 | -0.0489 |
| | (-0.51) | (-0.67) | (-0.65) | (-0.56) | (-0.68) | (-0.61) | (-0.69) | (-0.61) |
| satellite | 0.0445 | 0.0274 | 0.0336 | 0.0417 | 0.0266 | 0.0341 | 0.0279 | 0.0356 |
| | (0.66) | (0.41) | (0.51) | (0.63) | (0.40) | (0.51) | (0.42) | (0.54) |
| 3G | -0.0804 | -0.0878 | -0.0909 | -0.0823 | -0.0812 | -0.0889 | -0.0797 | -0.0873 |
| | (-1.29) | (-1.44) | (-1.47) | (-1.33) | (-1.37) | (-1.45) | (-1.35) | (-1.42) |
| USB | -0.0374 | -0.0486 | -0.0442 | -0.0418 | -0.0522 | -0.0456 | -0.0517 | -0.0451 |
| | (-0.75) | (-0.98) | (-0.88) | (-0.84) | (-1.04) | (-0.90) | (-1.03) | (-0.89) |
| frequency of meetings with friends | | 0.0517*** | 0.0552*** | | 0.0540*** | 0.0569*** | 0.0541*** | 0.0570*** |
| | | (8.58) | (9.27) | | (8.54) | (9.36) | (8.59) | (9.42) |
| social trust | | 0.200*** | | | 0.200*** | | 0.200*** | |
| | | (9.05) | | | (9.15) | | (9.17) | |
| social trust (wallet question) | | | 0.0873*** | | | 0.0866*** | | 0.0866*** |
| | | | (6.33) | | | (6.25) | | (6.24) |
| online networking | | | | -0.0412* | -0.0572** | -0.0492* | -0.0571** | -0.0490** |
| | | | | (-1.77) | (-2.51) | (-2.12) | (-2.50) | (-2.11) |
| real GDP per capita (thousands €2005) | | | | | | | -0.0784 | -0.0848 |
| | | | | | | | (-1.42) | (-1.63) |
| cut1 | -3.862*** | -3.514*** | -3.369*** | -3.906*** | -3.543*** | -3.401*** | -5.594*** | -5.618*** |
| | (-14.15) | (-13.85) | (-11.83) | (-14.23) | (-13.73) | (-11.87) | (-3.89) | (-4.15) |
| cut2 | -3.718*** | -3.367*** | -3.223*** | -3.762*** | -3.397*** | -3.255*** | -5.447*** | -5.472*** |
| | (-14.33) | (-14.02) | (-11.84) | (-14.42) | (-13.89) | (-11.89) | (-3.80) | (-4.06) |
| cut3 | -3.469*** | -3.115*** | -2.971*** | -3.513*** | -3.144*** | -3.003*** | -5.195*** | -5.220*** |
| | (-13.12) | (-12.56) | (-10.61) | (-13.22) | (-12.48) | (-10.66) | (-3.59) | (-3.83) |
| cut4 | -3.235*** | -2.879*** | -2.735*** | -3.279*** | -2.908*** | -2.768*** | -4.959*** | -4.985*** |
| | (-12.28) | (-11.66) | (-9.77) | (-12.34) | (-11.55) | (-9.80) | (-3.44) | (-3.67) |
| cut5 | -2.960*** | -2.602*** | -2.460*** | -3.005*** | -2.632*** | -2.491*** | -4.683*** | -4.707*** |
| | (-11.11) | (-10.38) | (-8.68) | (-11.15) | (-10.29) | (-8.74) | (-3.26) | (-3.48) |
| cut6 | -2.403*** | -2.039*** | -1.898*** | -2.447*** | -2.069*** | -1.929*** | -4.120*** | -4.146** |
| | (-9.22) | (-8.35) | (-6.84) | (-9.29) | (-8.30) | (-6.90) | (-2.87) | (-3.06) |
| cut7 | -1.796*** | -1.428*** | -1.288*** | -1.841*** | -1.455*** | -1.319*** | -3.506* | -3.536** |
| | (-6.78) | (-5.75) | (-4.59) | (-6.89) | (-5.74) | (-4.66) | (-2.45) | (-2.62) |
| cut8 | -0.920*** | -0.546* | -0.408 | -0.964*** | -0.573* | -0.438 | -2.624* | -2.655* |
| | (-3.50) | (-2.22) | (-1.47) | (-3.65) | (-2.28) | (-1.56) | (-1.84) | (-1.97) |
| cut9 | 0.121 | 0.500* | 0.636* | 0.0764 | 0.473* | 0.606* | -1.578 | -1.611 |
| | (0.46) | (2.04) | (2.30) | (0.29) | (1.90) | (2.17) | (-1.10) | (-1.19) |
| cut10 | 0.777** | 1.157*** | 1.293*** | 0.732** | 1.130*** | 1.263*** | -0.920 | -0.954 |
| | (2.99) | (4.77) | (4.71) | (2.79) | (4.56) | (4.55) | (-0.64) | (-0.71) |
| Observations | 16965 | 16921 | 16921 | 16921 | 16976 | 16965 | 16976 | 16965 |
| Pseudo R$^2$ | 0.026 | 0.029 | 0.028 | 0.026 | 0.030 | 0.028 | 0.030 | 0.028 |

*t* statistics in parentheses; * p < 0.1, ** p < 0.01. *** p < 0.001

**Table 3:** Life satisfaction and online networking: IV estimates using CMP

|  | without social capital | with social capital |
|---|---|---|
| life satisfaction | | |
| online networking | -0.0120 | -0.0408 |
|  | (-0.20) | (-0.54) |
| real GDP per capita (thousands €2005) | -0.00142 | -0.00116 |
|  | (-0.92) | (-0.74) |
| frequency of meetings with friends | | 0.0586*** |
|  | | (6.41) |
| social trust | | 0.222*** |
|  | | (12.41) |
| online networking | | |
| optic fiber (%) | 0.00501** | 0.00502** |
|  | (3.04) | (3.05) |
| broadband coverage | 0.00732*** | 0.00732*** |
|  | (3.56) | (3.55) |
| women | -0.125*** | -0.125*** |
|  | (-5.75) | (-5.73) |
| age | -0.0651*** | -0.0651*** |
|  | (-8.30) | (-8.29) |
| age squared/100 | 0.0297** | 0.0296** |
|  | (3.21) | (3.21) |
| N | 16921 | 16921 |
| F_stat | 14.95 | 14.94 |
| J_stat | 1247.8 | 1423.4 |
| chi2 | 2971.1 | 3196.7 |

$t$ statistics in parentheses; * $p < 0.1$, ** $p < 0.01$. *** $p < 0.001$
Regressions include socio-demographic and year controls.
Variables have been omitted for brevity and are available upon request.

**Table 4:** Life satisfaction and online networking: IV estimates using CMP

| | without social capital | with social capital |
|---|---|---|
| life satisfaction | | |
| online networking | -0.0120 | -0.0381 |
| | (-0.20) | (-0.50) |
| real GDP per capita (thousands €2005) | -0.00142 | -0.00127 |
| | (-0.92) | (-0.81) |
| frequency of meetings with friends | | 0.0606*** |
| | | (6.64) |
| social trust (wallet question) | | 0.0950*** |
| | | (7.87) |
| online networking | | |
| optic fiber (%) | 0.00501** | 0.00503** |
| | (3.04) | (3.05) |
| broadband coverage | 0.00732*** | 0.00732*** |
| | (3.56) | (3.56) |
| women | -0.125*** | -0.125*** |
| | (-5.75) | (-5.73) |
| age | -0.0651*** | -0.0651*** |
| | (-8.30) | (-8.27) |
| age squared/100 | 0.0297** | 0.0296** |
| | (3.21) | (3.20) |
| N | 16921 | 16921 |
| F_stat | 14.95 | 14.99 |
| J_stat | 1247.8 | 1374.0 |
| chi2 | 2971.1 | 3091.9 |

*t* statistics in parentheses; * $p < 0.1$, ** $p < 0.01$. *** $p < 0.001$
Regressions include socio-demographic and year controls.
Variables have been omitted for brevity and are available upon request.

**Table 5:** Indirect effects of the use of SNS on life satisfaction using SEM.

| | | |
|---|---|---|
| frequency of meetings with friends | | |
| online networking | 0.593*** | (31.86) |
| Constant | 5.010*** | (115.13) |
| life satisfaction | | |
| frequency of meetings with friends | 0.0821*** | (9.36) |
| social trust | 0.291*** | (8.40) |
| online networking | -0.0561** | (-2.03) |
| Constant | 7.022*** | (137.35) |
| social trust | | |
| online networking | -0.0271** | (-3.28) |
| Constant | 0.308*** | (22.69) |
| var(freq. of meetings with friends) | 1.418*** | (65.43) |
| var(life satisfaction) | 1.782*** | (52.71) |
| var(social trust) | 0.208*** | (39.92) |
| cov(freq. of meetings with friends, social trust) | 0.00249 | (0.50) |
| Observations | 16921 | |
| Indexes of goodness of fit | | |
| Chi-squared | (model vs. saturated) | 0.218 |
| Size of residuals | SRMR | 0.001 |
| Baseline comparison | | |
| CFI | 1.000 | Comparative fit index |
| TLI | 1.003 | Tucker-Lewis index |

$t$ statistics in parentheses; * $p < 0.1$, ** $p < 0.01$. *** $p < 0.001$

# Figures

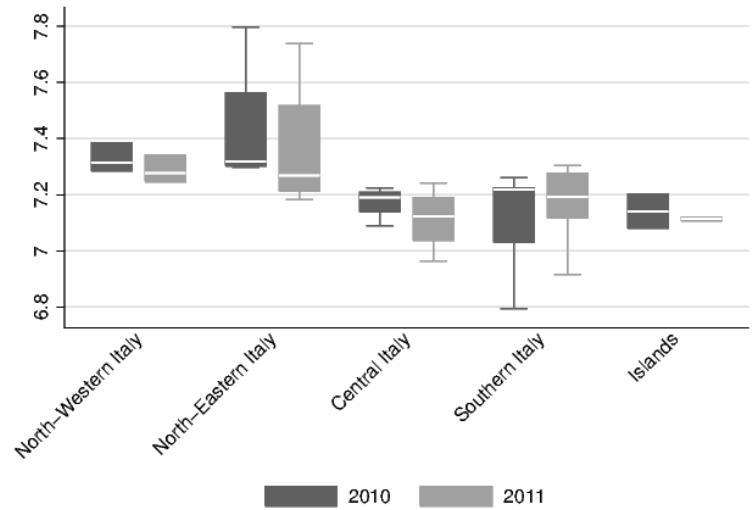

**Figure 1:** Distribution of average life satisfaction across regions and by year in Italy.

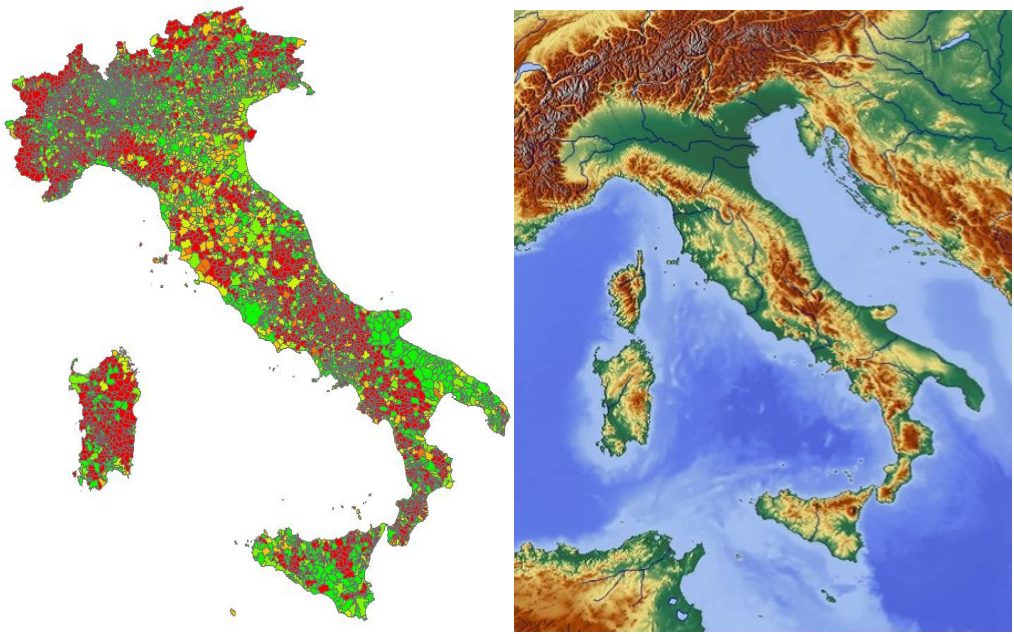

**Figure 2**: Percentage of the population covered by broadband in Italy (left) and topographic map of Italy (right). In the left figure, green areas have the best coverage and darker areas are those with the worst coverage. Sources: Between (2006), p. 17, and Wikimedia Commons.

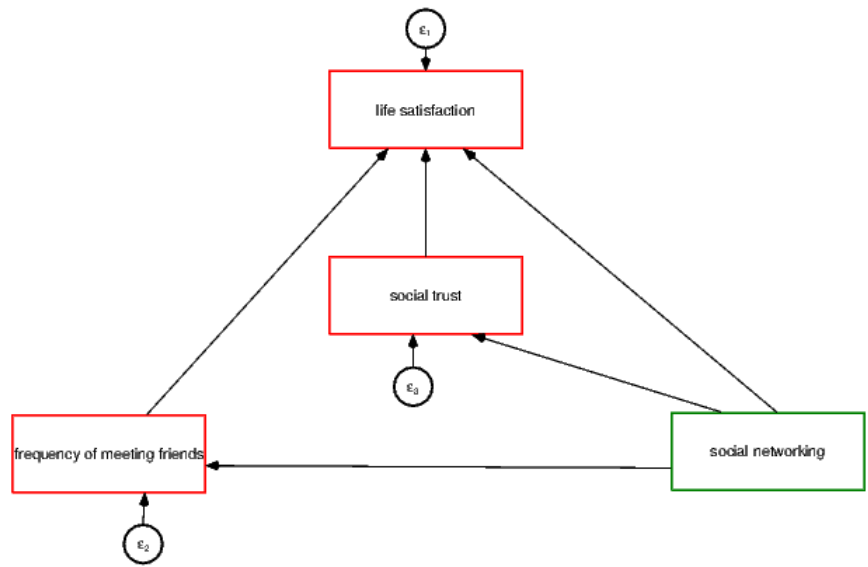

**Figure 3:** Structural equation model to estimate the direct and indirect effects of the use of SNS on subjective well-being.

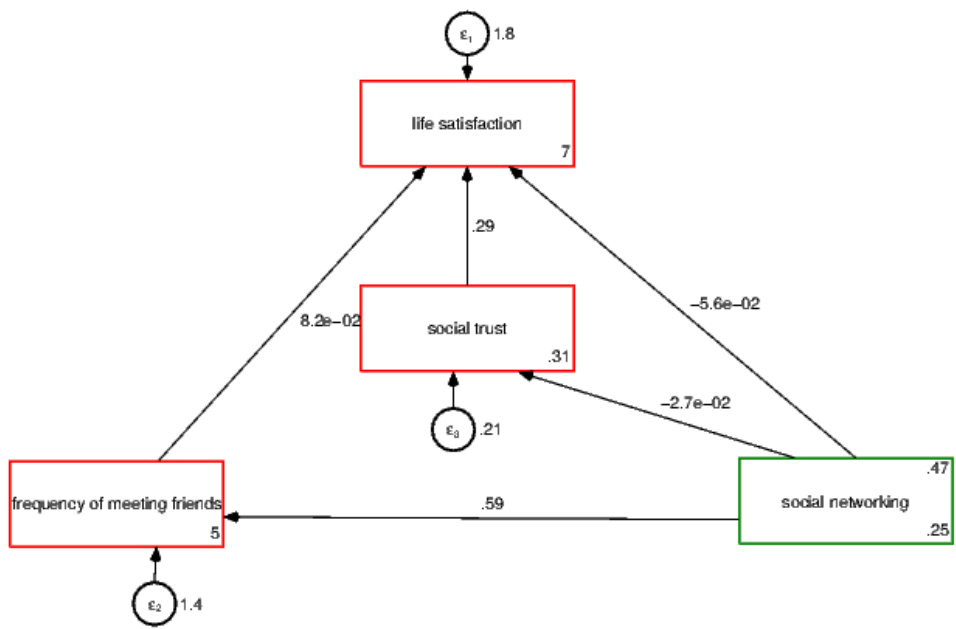

**Figure 4:** Direct and indirect effects of the use of SNSs on subjective well-being.